\documentclass[12pt,preprint,xdvi]{aastex}
\usepackage{amsmath} 
\usepackage{wrapfig} 
\usepackage{array} 
\usepackage{natbib}
\usepackage{brief_bib}

\renewcommand{\bibname}

\newcommand{\dmn}{{\cal D}}
\newcommand{\vol}{{\cal V}}
\newcommand\pvec{{\bf P}}
\newcommand\avec{{\bf A}}
\newcommand\bvec{{\bf B}}
\newcommand\nhat{{\bf \hat{n}}}
\newcommand\lhat{\textit{\textbf{\^l}}}
\newcommand\rvec{{\bf r}}
\newcommand\xvec{{\bf x}}
\newcommand\uvec{{\bf u}}

\newcommand\xhat{{\bf \hat{x}}}
\newcommand\yhat{{\bf \hat{y}}}
\newcommand\zhat{{\bf \hat{z}}}
\newcommand\uhat{{\bf \hat{u}}}
\newcommand\vhat{{\bf \hat{v}}}
\newcommand\phihat{{\bf \hat{\varphi}}}
\newcommand\psihat{{\bf \hat{\psi}}}

\newcommand\ex{e^{-\varpi^2/a^2}}

\begin{document}

\title{Additive Self Helicity as a Kink Mode Threshold}

\author{A. Malanushenko and D.W. Longcope}
\affil{Department of Physics, Montana State University\\
Bozeman, MT 59717, USA}

\and

\author{Y.~Fan and S.E.~Gibson} 
\affil{High Altitude Observatory, National Center for Atmospheric Research, P.O. Box 3000, Boulder, CO, 80307}

\begin{abstract}
In this paper we propose that additive self helicity, introduced by
\citet{Longcope2008e},  plays a role in the kink instability for
complex equilibria, similar to twist helicity for thin flux tubes
\citep{Hood1979,Berger1984}. We support this hypothesis by a
calculation of additive self helicity of a twisted flux tube from the
simulation of \citet{Fan2003}.  As more twist gets introduced, the
additive self helicity increases, and the kink instability of the tube
coincides with the drop of additive self helicity, after the latter
reaches the value of $H_A/\Phi^2\approx 1.5$ (where $\Phi$ is the flux
of  the tube and $H_A$ is additive self helicity). 

We compare additive
self helicity to twist for a thin sub-portion of the tube 
to illustrate that $H_A/\Phi^2$ is equal to the twist number, studied
by \citet{Berger1984}, when the thin flux tube  approximation is
applicable. We suggest, that the quantity $H_A/\Phi^2$ could be
treated as a generalization of a twist number, when thin flux tube
approximation is not applicable. A threshold on a generalized 
twist number might prove extremely useful studying complex equilibria,
just as twist number itself has proven useful studying idealized thin
flux tubes. We explicitly describe a numerical method for calculating
additive self helicity, which includes an algorithm for identifying a
domain occupied by a flux bundle and a method of calculating potential
magnetic  field confined to this domain. We also describe a numerical
method to calculate twist of  a thin flux tube, using a frame
parallelly transported along the  axis of the tube.
\end{abstract}

\section{Introduction}
According to a prevalent model coronal mass ejections (CMEs) are triggered by
current-driven magnetohydrodynamic (MHD) instability related to the 
external kink mode \citep{Hood1979,Torok2004, Rachmeler2009}.
The external kink mode, in its strictest form, is a helical deformation of
an initially symmetric,  cylindrical equilibrium, consisting of
helically twisted field lines.
The equilibrium is unstable to this instability if its field lines
twist about the axis by more than a critical angle, typically
close to $3\pi$ radians \citep{Hood1979,Baty2001}.  The helical 
deformation leads to an overall decrease in magnetic energy, since it
shortens many field lines even as it lengthens the axis.

Equilibria without symmetry can undergo an
analogous form of current-driven instability under which
global motion lowers the magnetic energy \citep{Bernstein1958,Newcomb1960}.  
Such an instability implies the existence of another equilibrium with
lower magnetic energy.  The spontaneous motion tends to deform the
unstable field into a state resembling the lower energy equilibrium.
Indeed, it is generally expected that there is at least one
minimum energy state from which deformation cannot lower the 
the magnetic energy without breaking magnetic field lines; its
energy is the absolute minimum under ideal motion.

Linear stability and instability are determined by the
energy change under infinitesimal motions. An
equilibrium will change energy only at the second order since first
order changes vanish as a requirement for force balance.
Ideal stability demands that no deformation decrease the energy at
second order, while instability will result if even one
energy-decreasing motion is possible.  The infinite variety of
possible motions make it impractical to establish stability in any
but the simplest and most symmetric equilibria.

Based on analogy to axisymmetric systems it is expected that general
equilibria, including those relevant to CMEs, are probably 
unstable when some portion of
their field lines are twisted about one another by more than some
critical angle.  This expectation was mentioned in a study by
\citet{Fan2003} of the evolution of a toroidal
flux rope into a pre-existing coronal arcade.  They solved
time-dependent equations of MHD in a three-dimensional, rectangular
domain.  Flux tube emergence was simulated by kinematically
introducing an isolated toroidal field through the lower
boundary.  The toroidal field was introduced beneath a pre-existing
arcade slowly enough that the coronal response never approached the
local Alfv\'en speed.  Fan and Gibson concluded that the system
underwent a current-driven instability after a critical amount of the
torus had been introduced.  They bolstered this claim by performing an
auxiliary run where the kinematic emergence was halted and the system
allowed to evolve freely; it settled into an equilibrium.

While twist angle has proven useful in a few cases, it is difficult to
demonstrate its utility as a threshold in general, asymmetric equilibria.
Indeed, in any but a few very symmetric cases there is no
simple, obvious way to define the angle by which the field lines wrap
about one another.  The local rate of twist is given by the current
density, which is after all the source of free energy powering the
instability.  On the other hand, excessive local current density is
not sufficient to drive instability.  This fact is illustrated by
numerous examples of discontinuous field which are minimum energy
states.

It has been suggested that a threshold exists, in general equilibria,
for some global quantity such as free magnetic energy or helicity
\citep{Zhang2006, Low1994}.  If this is the case then we expect the instability to
lower the value of this global quantity so that it falls below the
threshold value in the lower-energy, stable equilibrium.
Magnetic helicity is a logical candidate to play
this role since it is proportional to total twist angle in cylindrical
fields.  Relative helicity in particular is a proxy for currents.
Helicity is, however, conserved under ideal motion and therefore
will not be reduced to a sub-threshold value by an ideal instability.

The total helicity of a thin, isolated flux tube can be written as a
sum of two terms called twist and writhe \citep{Berger1984,Moffatt1992},
$$H=H_{T}+H_{W}.$$
The writhe depends on the
configuration of the tube's axis while the twist depends on the
wrapping of field lines about one another. A cylindrical tube has a
perfectly straight axis and therefore zero writhe helicity.  Any ideal
motion which helically deforms the entire flux tube will increase the
magnitude of the writhe helicity.  Since the motion 
preserves total helicity the change in writhe must be accompanied by an
offsetting change in twist helicity.  If the writhe has the same sign
as the initial twist, then the motion will decrease the twist
helicity.  In cases where the magnetic energy depends mostly on twist,
this motion will decrease the magnetic energy \citep{Linton2002}.  The
straight equilibrium is therefore unstable to an external kink mode.

Topologically, the foregoing properties of magnetic field lines could be 
compared to the properties of thin closed ribbons. One may introduce 
\textit{twist number}, \textit{writhe number} and their combination, called \textit{linkage} number, is 
a preserved quantity in the absence of reconnection
\citep{Berger1984,Moffatt1992},
$$L=Tw+Wr.$$

By analogy to the case of a thin isolated flux tube we consider the
twist helicity, rather than the total helicity, to be the most likely
candidate for a stability threshold.  Indeed, within a thin flux tube
it is possible to derive a net twist angle among field lines and
$H_T=\Phi^2Tw=\Phi^2\Delta\theta/2\pi$, where $\Phi$ is the total magnetic 
flux through a cross-section of the tube and $\Delta\theta$ is the net twist angle.

Twist and writhe are, however, defined only in cases of thin, isolated
magnetic flux tubes, and can no more set the threshold we seek
than the net twist angle can.

Recently \citet{Longcope2008e} introduced two generalizations of relative
helicity applicable to arbitrary sub-volumes of a magnetic field.
They termed both generalized self-helicity, and the two differed only
by the reference field used in their computation.  The one called 
{\em additive self-helicity} (that we denote $H_A$) uses a reference 
field confined to the same sub-volume as the original field, and can 
be interpreted as a generalization of the twist helicity to arbitrary 
magnetic fields. The additive self-helicity of a thin, isolated flux 
tube is exactly the twist helicity.

Since the additive self-helicity can be computed for arbitrary
magnetic fields we propose that it (normalized by the squared flux) is the 
quantity to which current-driven instability sets an upper limit, which 
could be considered a generalized twist number: 
\begin{equation}
Tw_{(gen)}=H_A/\Phi^2.
\label{gen_twist}
\end{equation}

The paper is organized as follows. In Section 2, we describe a method 
for calculating additive self helicity and $Tw_{(gen)}$ numerically. 
There are two large and nontrivial parts of this calculation, that we describe 
in 2.1 and 2.2: locating a domain containing a given flux bundle 
and constructing a potential field 
in this domain by Jacobi relaxation. In Section 3, we apply the method
to a simulation  to support our hypothesis, the emerging twisted flux
tube from \citet{Fan2003}. In 3.1 we briefly describe this
simulation, and then in 3.2 we show different embedded domains defined
by different subportions of the footpoints. In 3.2
we describe, how the twist of \citet{Berger1984} could be calculated
for those of the domains for which thin flux tube approximation is
applicable. 
In Section 4 we present the evolution of additive self
helicity, unconfined self-helicity, twist (for ``thin''  domains) and
the integrated helicity flux in the simulation. 
We demonstrate that $Tw_{(gen)}$
increases corresponding to helicity flux, that it drops after it
reaches a certain value (about $1.5$) and that this drop coincides
with the rapid expansion of the tube due to the kink 
instability. We also demonstrate that the unconfined self helicity
grows only when  helicity flux is nonzero and that it stays constant
when kink instability happens.  We also show that $Tw_{(gen)}$
corresponds to $Tw$ when thin flux tube approximation  is applicable.

\clearpage
\section{Numerical Solutions}
 
The object of study is a magnetic field $\bvec\left(\rvec\right)$
defined in a domain $\dmn$, $\rvec\in\dmn$, that lies on and above the
photosphere, $z\geq 0$. By domain we understand a volume that
encloses the field: $\bvec\cdot\nhat=0$ on all boundaries,
$\partial\dmn$, except at the photosphere, where
$\bvec\cdot\nhat=B_z(x, y, z=0)$. An example of such a volume is the
coronal part of an $\Omega$-shaped loop. The self-helicity is given
by 
\begin{equation}
 H_A\left(\bvec, \pvec\left(\dmn\right), \dmn\right)=\int\limits_{\dmn}{\left(\bvec-\pvec\right)\cdot\left(\avec+\avec_P\right)dV},
\end{equation}
as defined in \citet{Longcope2008e}. Here $\pvec$ is the potential
magnetic field, whose normal component matches the normal component of
$\bvec$ on the boundary $\partial\dmn$, 
\begin{equation}
 \left.\pvec\cdot\nhat\right|_{\partial\dmn}=\bvec\cdot\nhat|_{\partial\dmn},
\end{equation}
$\avec$ and $\avec_P$ are the vector potentials of $\bvec$ and $\pvec$ respectively (as discussed in \citet{Finn1985}, helicity, defined this way is gauge-independent).	\\

Once the self-helicity is known, the twist is given by eq.\
(\ref{gen_twist}) with $\Phi$ being the total signed flux of the
footpoints of the configuration: 
\begin{equation}
 \Phi=\int\limits_{z=0,B_z\geq 0}{B_zdxdy}=-\int\limits_{z=0,B_z\leq 0}{B_zdxdy}.
 \label{flux}
\end{equation}
In the next two sections we discuss methods of numerically obtaining
$\dmn$, from given footpoints, and $\pvec$.

 \subsection{Finding the domain.} \label{findingthedomain}

 In order to describe the domain on a grid we introduce the support function:
$$\Theta(\rvec)=\left\{ \begin{aligned}
1&\mbox{, if }\rvec\in \dmn\\
0&\mbox{, if }\rvec\notin \dmn\mbox{.}\\
\end{aligned}\right.
$$
This is a function of the given magnetic field $\bvec$ and some
photospheric area, called the boundary mask. By definition, every
field line, initiated at any point on the boundary mask and having the
other footpoint somewhere within the mask, is completely inside the
domain $\dmn$. If the field line traced in both directions from some
coronal point ends within the photospheric mask, then this point also
belongs to the domain. In numerical computations we replace ``point''
with a small finite volume, voxel $v_{ijk}$ (3-dimensional pixel). We
define a voxel to be inside $\dmn$ (equivalent to saying
$\Theta\left(v_{ijk}\right)=1$), if there is \textit{at least one}
point inside it that belongs to $\dmn$.\\
 \begin{figure}[!hc]
 \begin{center}
    \includegraphics[width=6cm]{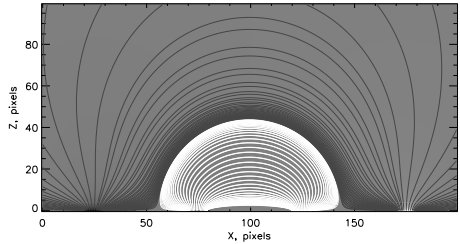} 
 \end{center} 
  \caption{\small{An example of what we call a domain. Here the field from four photospheric sources is computed on a half-space. Two possible domains are shown in two different colours.}}
  \label{sample_domains} 
 \end{figure}

The simplest method of constructing the support function would be to
trace a field line in both direction from every voxel of the
computational grid, set $\Theta=1$ in the voxel if the footpoints
both terminate in pixels from the boundary regions, and set $\Theta=0$
otherwise. This, however, is a very time-consuming algorithm,
especially for a large arrays of data.
Instead we use an algorithm which reduces the computational time by tracing
field lines from a subset of voxels. It works by progressively adding
voxels to $\Theta$ adjacent to those already known to belong to
$\dmn$. \\

We add a voxel centered at $r_{i,j,k}$ to the domain under two
different circumstances. 1. A field line initialized somewhere within
the volume of the voxel $v_{i,j,k}$, centered at $r_{i,j,k}$, is found
to have both footpoints within the boundary mask. 2. A field line
initiated in some other voxel, and determined to belong to $\dmn$,
passes through some portion of the volume $v_{i,j,k}$.\\

Initially, the domain consists only of footpoint voxels, so the
initial step is to trace field lines \textit{initiated at the
footpoints}, assuming, that at least some of these lines will lie in
the domain. \\

\begin{figure}[!hc]\begin{center}
  \includegraphics[angle=-90]{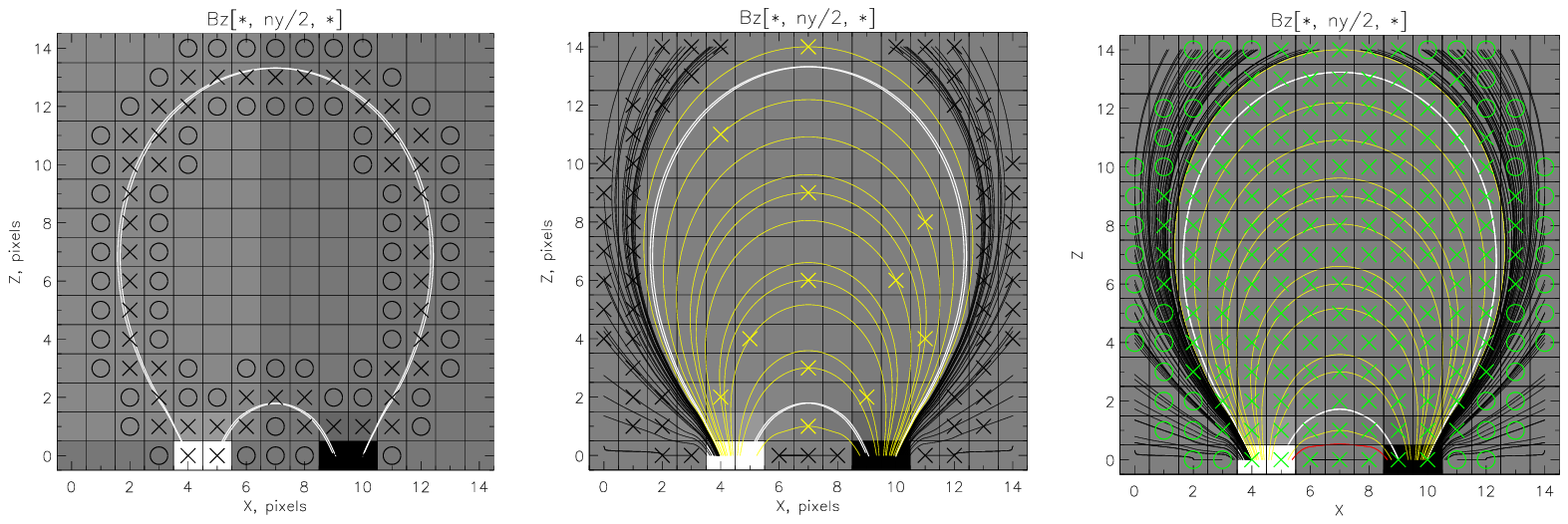} 
  \end{center}
  \caption{\small{\textit{(left)} --- The first iteration of the
iterative search: from the initially selected voxels (crosses), check
those surrounding (circles) for membership in the domain. Repeat until
no ``surrounding'' voxels can be added to the
domain. \textit{(middle)} --- The voxels, checked on \textit{all}
iterations in the middle plane. For every field line, a cross shows
where it was initialized. Yellow are ``accepted'' lines (and thus
\textit{all} voxels that contain them are ``accepted'') and black are
``not accepted'' lines (and thus \textit{only} voxels where these
lines were initializes from are ``not accepted''). \textit{(right)} -- The end result. The green crosses mark voxels that are found to belong to the domain and the green circles are the neighbourhood of the domain. White (initial), yellow (iterative) and red (final) field lines are traced and found to be in the domain; black lines are found to be not in the domain. Note that the domain is ``covered'' by much fewer lines than an exhaustive search would do.}}
  \label{secondary} 
 \end{figure}

We illustrate the method on a simplistic case of a potential magnetic field, confined to a half-space, with $B_z=0$ everywhere at the photosphere, except at four pixels, as shown on Fig.~\ref{secondary}. We have computed the magnetic field inside a small box of $15\times15\times15$ pixels, centered around the photospheric sources. The boundary mask consists of these four voxels at the photosphere with non-zero vertical magnetic field. In this simplistic example the initial guess would be four field lines, initiated at four footpoint voxels, as shown on Fig.~\ref{secondary}, left (note that in this particular example a field line, initiated at one voxel, ends at another voxel within the mask and thus is the same as the field line, initiated at that another voxel, so these four initial guesses are really two, not four field lines). The voxels of the initial guess are shown with crosses.\\

In an algorithm, this would be the first step:
\begin{itemize}
	\item[Step 1:]{Make the initial guess: trace field lines from the footpoints.}
\end{itemize}
As soon as an initial step is made, the next step is to assume, that
the \textit{immediate neighbourhood} of voxels known to be in $\dmn$
are likely to be also in the domain. Thus, in the next (iterative)
search the following steps are performed:
\begin{itemize}
	\item[Step 2:]{Locate voxels on the boundary of the current domain.}
	\item[Step 3:]{For every voxel on the boundary: trace a field line and check whether it is in the domain. 
                \begin{itemize}
	               \item[If yes:]{Add the voxel to the domain. Add
all voxels along the line to the domain. Exclude them from the
boundary (there is no need to check them again).}
	               \item[If no:]{Mark the voxel as ``questionable''. (If there is a field line, which passes through the voxel and does not belong to the domain, then at least part of the voxel is outside of the domain. Since its immediate neighbourhood is in the domain, then it is possible that part of it is also in the domain.)}      
	              \end{itemize}
	             }
	\item[Loop:]{Repeat steps 2-3 until \textit{all} the voxels in the boundary are ``questionable'' and no new voxels are added.}
\end{itemize}

When the iterative search does not find any new voxels, we make the
final check of the boundary voxels. The idea is to trace field lines
from all corners of such ``questionable'' voxels to see, which corners
(and thus which part of a voxel) belongs to the domain. We consider
this to be optional check, which may improve the precision of the
definition of the domain by at most one layer of voxels. \\  

This last search may also give information about the normal to the
domain surface. If it is known that some corners of a voxel are
in the domain and some are not, it is possible to approximate the
boundary as a plane separating those two groups of corners. \\
\begin{itemize}
	\item[Step 4, optional:]{For each voxel, marked previously as ``questionable'', check the corners (by tracing field lines) to see which of them are in the domain and which aren't. Keep this information.} 
\end{itemize}

\clearpage
 \subsection{Constructing the Confined Potential Field $\pvec$}

Once the domain has been determined, the next step is to construct the
potential magnetic field confined to it. We use a common relaxation
method on a staggered grid in orded to account for the complex
boundaries of $\dmn$. \\

We introduce a scalar potential $\bvec_p=\nabla\chi$ and look for the solution of the Laplace's equation for $\chi$
$$\nabla\cdot\bvec_P=\nabla^2\chi=0.$$ 
By the definition of $\dmn$, field lines never cross $\partial\dmn$,
except at the lower boundary, $z=0$. Thus, boundary conditions for
$\bvec_p$ could be written as:
$\left.\bvec_p\cdot\nhat\right|_{\partial\dmn, z\neq 0}=0$ and
$\left.\bvec_p\cdot\nhat\right|_{\partial\dmn, z=0}=\bvec_z(x,
y)$. This is equivalent to Neumann boundary conditions for $\chi$:

\begin{equation}
 \begin{array}{rcl}
  \left.\frac{\partial\chi}{\partial n}\right|_{\partial\dmn, z\neq 0}&=&0,\\
  \left.\frac{\partial\chi}{\partial z}\right|_{\partial\dmn, z=0}&=&B_z(x, y).
 \end{array}
\label{bc_eqn}
\end{equation}

\textbf{The Algorithm for the Relaxation Method} \\

We use the Jacobi iterative method
\citep[see, for example][]{LeVequeFDM} to solve for the potential
field. Here we briefly summarize the algorithm and further explain in
details. The $n+1$-th iteration is
\begin{enumerate}
	\item{$\forall\rvec\in\dmn$: calculate a new iteration
$\chi^{[n+1]}$ as a solution of the equation
$\chi^{[n+1]}-\chi^{[n]}=Kh^2\nabla^2\chi^{[n]}$, where $h$ is the
grid spacing. The Laplacian $\nabla^2\chi^{[n]}\left(\rvec\right)$,
found ausing standard finite difference methods, is equivalent to an
average over some stencil of neighbouring points minus the central
value; $K$ is a constant that depends on the exact shape of the
stencil.}
	\item{$\forall\rvec_b\in\partial\dmn$: set
$\chi^{[n+1]}\left(\rvec_b\right)$ so as to satisfy boundary
conditions (BCs).}
	\item{Repeat steps 1--2, until the difference between
$\chi^{[n]}\left(\rvec\right)$ and $\chi^{[n+1]}\left(\rvec\right)$
is sufficiently small in some sense (namely, until
$||\chi^{[n+1]}-\chi^{[n]}||<\epsilon$, where $\epsilon$ is
pre-defined small number).}
\end{enumerate}

\textbf{Staggered Mesh}\\

The functions $B_x(x, y, z)$, $B_y(x, y, z)$ and $B_z(x, y, z)$ are defined on the same mesh points $(x_i, y_j, z_k)$. If we are interested in finding $\chi(x, y, z)$, so that $B_x=\frac{\partial\chi}{\partial x}$,  $B_y=\frac{\partial\chi}{\partial y}$ and $B_z=\frac{\partial\chi}{\partial z}$, it advantageous to define $\chi$ \textit{in between} the original mesh points 
and calculate the derivatives using finite difference as following: 

$$\begin{array}{lcr}
B_x(x_i, y_j, z_k)&=&\cfrac{\chi(x_{i+1/2}, y_j, z_k)-\chi(x_{i-1/2}, y_j, z_k)}{x_{i+1/2}-x_{i-1/2}},\\
\end{array}, $$
and so on for $B_y$ and $B_z$. $\chi$, then, would only be defined in the middle of \textit{the faces} of cubic voxels, i.e., at points $(i\pm 1/2, j, k)$, $(i, j\pm 1/2)$ and $(i, j, k\pm 1/2)$.\\
Such a mesh, called a \textit{``cartesian staggered mesh''}, is known
to have better numerical properties, such as immunity from decoupling
of variables and having a smaller numeric dispersion 
\citep[][see, for example]{perot}. 
\\


The finite difference approximation of a Laplacian at one point can be
interpreted as a weighted average over a stencil of several points
minus the value at that point. For example, in the 2D case the second
order approximation to $\nabla^2 \chi(x, y)$ on a uniform Cartesian grid
at the point $\left(x_i, y_j\right)$ could be computed over a 5-point
stencil: 
$$\nabla^2\chi\left(x_i, y_j\right)\approx\frac{1}{h^2}\left(\chi\left(x_{i-1}, y_j\right)+\chi\left(x_{i+1}, y_j\right)+\chi\left(x_i, y_{j-1}\right)+\chi\left(x_i, y_{j+1}\right)-4\chi\left(x_i, y_j\right)\right)$$ 
(here $h$ is the spacing of the grid). It could be rewritten as 
$$\chi\left(x_i, y_j\right)\approx\frac{1}{4}\left(\chi\left(x_{i-1}, y_j\right)+\chi\left(x_{i+1}, y_j\right)+\chi\left(x_i, y_{j-1}\right)+\chi\left(x_i, y_{j+1}\right)\right)-\frac{h^2}{4}\nabla^2\chi\left(x_i, y_j\right).$$ 
The Jacobi method uses this equation to iteratively update the value at the point, constantly assuming $\nabla^2\chi=0$. In the case of the 5-points stencil the updated value would be 
$$\chi^{[n+1]}\left(x_i, y_j\right)=\frac{1}{4}\left(\chi^{[n]}\left(x_{i-1}, y_j\right)+\chi^{[n]}\left(x_{i+1}, y_j\right)+\chi^{[n]}\left(x_i, y_{j-1}\right)+\chi^{[n]}\left(x_i, y_{j+1}\right)\right).$$

\begin{figure}[!hc]
 \begin{center}
    \includegraphics[width=12cm]{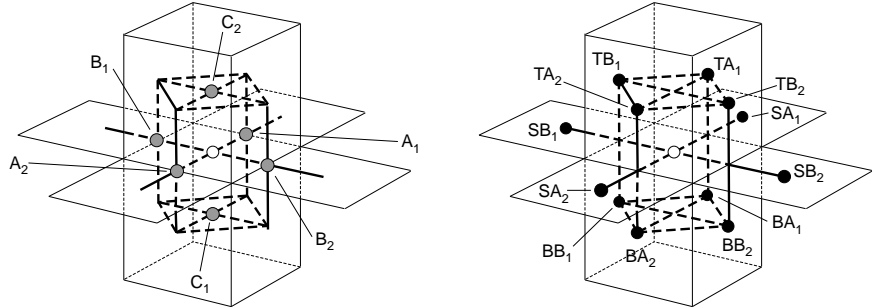} 
 \end{center} 
  \caption{\small{The averaging kernel for the laplace's equation on 3D staggered mesh (right) and the motivation for it (left). For example, the stencil for a face with normal vector $\zhat$ would include five ``$\zhat$ faces'' (including itself), four ``$\xhat$ faces'' and four ``$\yhat$ faces'' (two of each above and below). }}
  \label{stagg_kernel} 
 \end{figure}

In our case of a 3D staggered mesh, choosing a stencil becomes more
complicated. We propose a 13-point scheme, shown on the right of
Fig.~\ref{stagg_kernel} (black dots). To motivate this stencil,
we derive it from the ``unstaggered'' one (Fig.~\ref{stagg_kernel},
left, gray dots). In an ``unstaggered'' finite differencing
scheme the $[n+1]$-th iteration in Jacobi method would be expressed as
$$6\chi^{[n+1]}(O)=\chi^{[n]}(A_1)+\chi^{[n]}(A_2)+\chi^{[n]}(B_1)+\chi^{[n]}(B_2)+\chi^{[n]}(C_1)+\chi^{[n]}(C_2).$$
But for the staggered mesh $\chi$ is undefined at these nodes. This
can be resolved by setting $\chi$ at each ``gray'' point to be equal
to the \textit{average} of its 4 closest neighbours,
$$\begin{array}{rcl}
\chi^{[n]}(A_1)&=&\frac{1}{4}\left[\chi^{[n]}(SA_1)+\chi^{[n]}(TA_1)+\chi^{[n]}(SA_1)+\chi^{[n]}(O)\right],\\
\chi^{[n]}(B_1)&=&\frac{1}{4}\left[\chi^{[n]}(SB_1)+\chi^{[n]}(TB_1)+\chi^{[n]}(SB_1)+\chi^{[n]}(O)\right],\\
\chi^{[n]}(C_1)&=&\frac{1}{4}\left[\chi^{[n]}(TA_1)+\chi^{[n]}(TA_2)+\chi^{[n]}(TB_1)+\chi^{[n]}(TB_2)\right]
\end{array}$$
and so on. Then we may substitute this in the original expression and get: 
$$\begin{array}{rcl} 6\chi^{[n+1]}(O)&=&2\times\frac{1}{4}\left[\chi^{[n]}(TA_1)+\chi^{[n]}(TA_2)+\chi^{[n]}(BA_1)+\chi^{[n]}(BA_2)\right]+\\
&+&2\times\frac{1}{4}\left[\chi^{[n]}(TB_1)+\chi^{[n]}(TB_2)+\chi^{[n]}(BB_1)+\chi^{[n]}(BB_2)\right]+\\ &+&\frac{1}{4}\left[\chi^{[n]}(SA_1)+\chi^{[n]}(SA_2)+\chi^{[n]}(SB_1)+\chi^{[n]}(SB_2)\right]+\\
&+&4\times\frac{1}{4}\chi^{[n]}(O), 
\end{array}$$ 
which is eqivalent to
$$\begin{array}{rcl}
 \chi^{[n+1]}(O)&=&\frac{1}{12}\left[\chi^{[n]}(TA_1)+\chi^{[n]}(TA_2)+\chi^{[n]}(BA_1)+\chi^{[n]}(BA_2)\right]+ \\
 &+&\frac{1}{12}\left[\chi^{[n]}(TB_1)+\chi^{[n]}(TB_2)+\chi^{[n]}(BB_1)+\chi^{[n]}(BB_2)\right]+\\         &+&\frac{1}{24}\left[\chi^{[n]}(SA_1)+\chi^{[n]}(SA_2)+\chi^{[n]}(SB_1)+\chi^{[n]}(SB_2)\right]+\\
&+&\frac{1}{6}\chi^{[n]}(O). 
\end{array}$$ 
With these weights th ``farthest'' nodes $S[AB]_{[12]}$ have half the
influence on the laplacian, of the ``closer'' nodes. Note also, that
the sum of the weights is one. \\

\textbf{Boundary Conditions} \\


Boundary conditions (given by Eqn.~\ref{bc_eqn}) in the staggered mesh
is particularly easy if one assumes that the boundary surface passes
\textit{inside} of boundary voxels, rather than on their
sides. Suppose, for example, that the boundary plane normal to $\zhat$
passes through the center of the voxel $v_{ijk}$.
Then the BC for this voxel would be that $B_z\left(i,
j, k\right)=0$, or simply $\chi\left(i, j,
k+\frac{1}{2}\right)=\chi\left(i, j, k-\frac{1}{2}\right)$.\\

To motivate such choice of the boundary, we note that boundary voxels,
by definition, are the voxels \textit{part of which is inside of
$\dmn$ while part is outside}. Such a conclusion is made about voxels,
some of whose corners are inside of $\dmn$, and some of the corners
are outside of $\dmn$ (this information about the domain is obtained
in the step 4 of the algorithm, described in
section~\ref{findingthedomain}). We approximate the boundary inside of
each boundary voxel as a plane, that passes through the center of the
voxel and that separates its ``exterior'' part from its ``interior''
part. Such approximation will err by no more that
$1/\sqrt{2}$ voxel's length off the real location of the boundary.
We also find it easier to work in terms of faces
rather than corners, since this is where $\chi$ is defined. (We say,
that a face is ``exterior'' to the domain if more than two of its
corners are not in the domain, i.e., for a voxel, we say, that if only
one corner or only one edge are ``exterior'', we do not consider it a
subject to BC's). \\

\begin{figure}[!hc]
 \begin{center}
    \includegraphics[width=10cm]{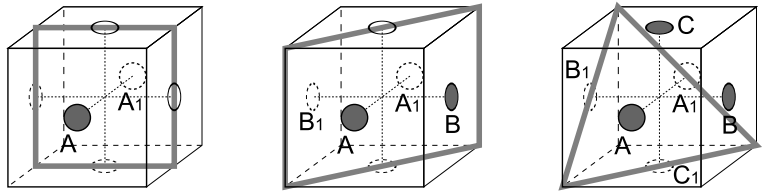} 
 \end{center} 
  \caption{\small{Different ways to approximate the boundary surface inside of a boundary voxel, depending on which portion of the voxel is found to be outside of the domain. White dots are the centers of the ``interior'' faces, gray dots are the centers of the ``exterior'' faces (see explanation in the text), the thick plane is the proposed approximation of the boundary surface $\partial\dmn$.}}
  \label{bc} 
 \end{figure}

There are several ways to orient such a boundary plane inside a voxel,
based on the behaviour of the boundary in the immediate surrounding of
the voxel.
\begin{enumerate}
	\item{The voxel has only one face outside of the domain. Then we consider the boundary parallel to that face of the voxel (see Fig.~\ref{bc}, left). If, say, the boundary is parallel to the face between faces $A$ and $A_1$ (see Fig.~\ref{bc}, bottom left), then the normal field to the boundary is $\bvec\cdot\bf\widehat{AA}_1$ (hereafter $\bf\widehat{AA}_1$ denotes a unit vector along the line from $A$ to $A_1$, which might be $\pm\xhat$, $\pm\yhat$ or $\pm\zhat$), and BC would be formulated as \\ 
$\begin{array}{lcccccc}
  \chi_A&=&1\cdot\chi_{A'}&+&0\cdot\chi_{B'}&+&0\cdot\chi_{C'}
 \end{array}$.} 
	\item{The voxel has two adjacent faces outside of the domain. Then we approximate the boundary as a plane, that cuts off these two faces, as shown on Fig.~\ref{bc},  middle. If faces $A$ and $B$ are outside and faces $A_1$ and $B_1$ are inside of the domain, then we consider the normal field to be $\bvec\cdot\frac{1}{\sqrt{2}}\left(\bf{\widehat{AA}_1}+\bf{\widehat{BB}_1}\right)$ and set BC's as \\
	$\begin{array}{lcccccc}
\chi_A&=&0\cdot\chi_{A'}&+&1\cdot\chi_{B'}&+&0\cdot\chi_{C'},\\
\chi_B&=&1\cdot\chi_{A'}&+&0\cdot\chi_{B'}&+&0\cdot\chi_{C'}.\\
\end{array}$ }
  \item{Similarly, if three mutually adjacent faces of the voxel are outside of the domain (and three others are inside), as shown on Fig.~\ref{bc}, right, then, analogously, we assume that the normal field is $\bvec\cdot\frac{1}{\sqrt{3}}\left(\bf{\widehat{AA}_1}+\bf{\widehat{BB}_1}+\bf{\widehat{CC}_1}\right)$ and BC's could be set in the following way: \\
  $\begin{array}{lcccccc}
\chi_A&=&0\cdot\chi_{A'}&+&\frac{1}{2}\cdot\chi_{B'}&+&\frac{1}{2}\cdot\chi_{C'},\\
\chi_B&=&\frac{1}{2}\cdot\chi_{A'}&+&0\cdot\chi_{B'}&+&\frac{1}{2}\cdot\chi_{C'},\\
\chi_C&=&\frac{1}{2}\cdot\chi_{A'}&+&\frac{1}{2}\cdot\chi_{B'}&+&0\cdot\chi_{C'}.\\
\end{array}$\\ (Note that in this case there are really three variables and one equation to satisfy; thus, there are different solutions to $\chi$. But each of those solutions would be valid, as long as it satisfies $\bvec\cdot\nhat=0$.)}
  \item{``Everything else'': the voxel has three or more non-adjacent faces that are outside of the domain, but still is on the boundary. It is considered an extraneous voxel and is removed from the boundary.}
\end{enumerate}

\clearpage
\section{The Experiment}
 
The method described above was tested on a simple quadrupole example,
and the values of self-helicity it gives are in a good agreement with
theoretical predictions \citep{Longcope2008e}. That work, however,
does not consider any sort of stable equilibrium and does not study
any kinking instability thresholds, similar to those developed in
Hood~\&~Priest, 1981\nocite{Hood1981}. \\ 
 
The objective of the current work is to test whether the parameter
$H_A/\Phi^2$ behaves like a total twist in the sense that
it has a critical value above which a system is unstable to a global
disruption. To do so, we use the numerical simulation of kink
instability in an emerging flux tube from Fan \& Gibson, 2003
\nocite{Fan2003}. \\

\subsection{Simulation Data}
 
\begin{figure}[!hc]
 \begin{center}
    \includegraphics[width=4cm]{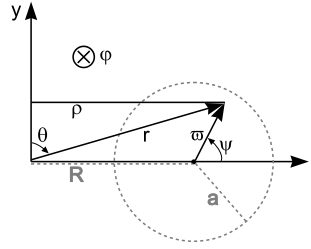} 
 \end{center} 
  \caption{\small{$\bvec$ is set in spherical coordinates $\left(r, \theta, \phi\right)$ with the polar axis directed along $\yhat$. We will  mainly use different coordinates, namely, $\left(\varpi, \psi, \phi\right)$. $R$ is the major radius of the torus, $a$ is the minor one, $\rho=r\sin\left(\theta\right)$ is the distance from the y axis.}}
  \label{toroidal_coord} 
 \end{figure}

The initial configuration is a linear arcade above the
photosphere, into which a thick, non-force-free torus was emerged. Inside the 
torus the field lines wind around its minor axis and the
field magnitude drops with distance from the minor axis. The exact
shape of the magnetic field, in the coordinates shown on
Fig.~\ref{toroidal_coord}, is the following:  
\begin{equation}
 \bvec_0=B_\psi\psihat+B_\varphi\phihat=B_t\ex\left(q\frac{\varpi}{\rho}\psihat+\frac{a}{\rho}\phihat\right),
 \label{b0}
\end{equation}
where $a=0.1L$ is the minor radius, $R=0.375L$ is the major radius,
$q=-1$, $B_t=9B_0$, $L$ is the length scale of the domain (further in
our calculations $L=1$), $B_0$ is the characteristic strength of the
photospheric arcade the torus is emerging into, and the time is given
in the units of Alfven time, $\tau_A=L/v_A$. The field
strength drops as $\ex$ with $\varpi$ being the distance from the
minor axis. At $\varpi=3a$ magnetic field was artificially set to 0.\\

 The torus is ``emerged'' from underneath the photosphere with a
constant speed. There is a mass flow across the photosphere in the area,
and the emerging tube is driven into the domain by an electric field at 
the boundary. This is made in the following way: for each time step 
(starting at t=0 and until the axis of the torus has emerged, t=54) the 
vertical photospheric field is set to that from the appropriate slice 
of the torus's field. Dynamical equations are then solved in order for
the field above $z=0$ to relax, so that at every time step the resulting 
configuration is a force-free equilibrium. The unsigned photospheric flux 
as a function of time is shown on Fig.~\ref{flux}. 

\begin{figure}[!hc]
 \begin{center}
   \includegraphics{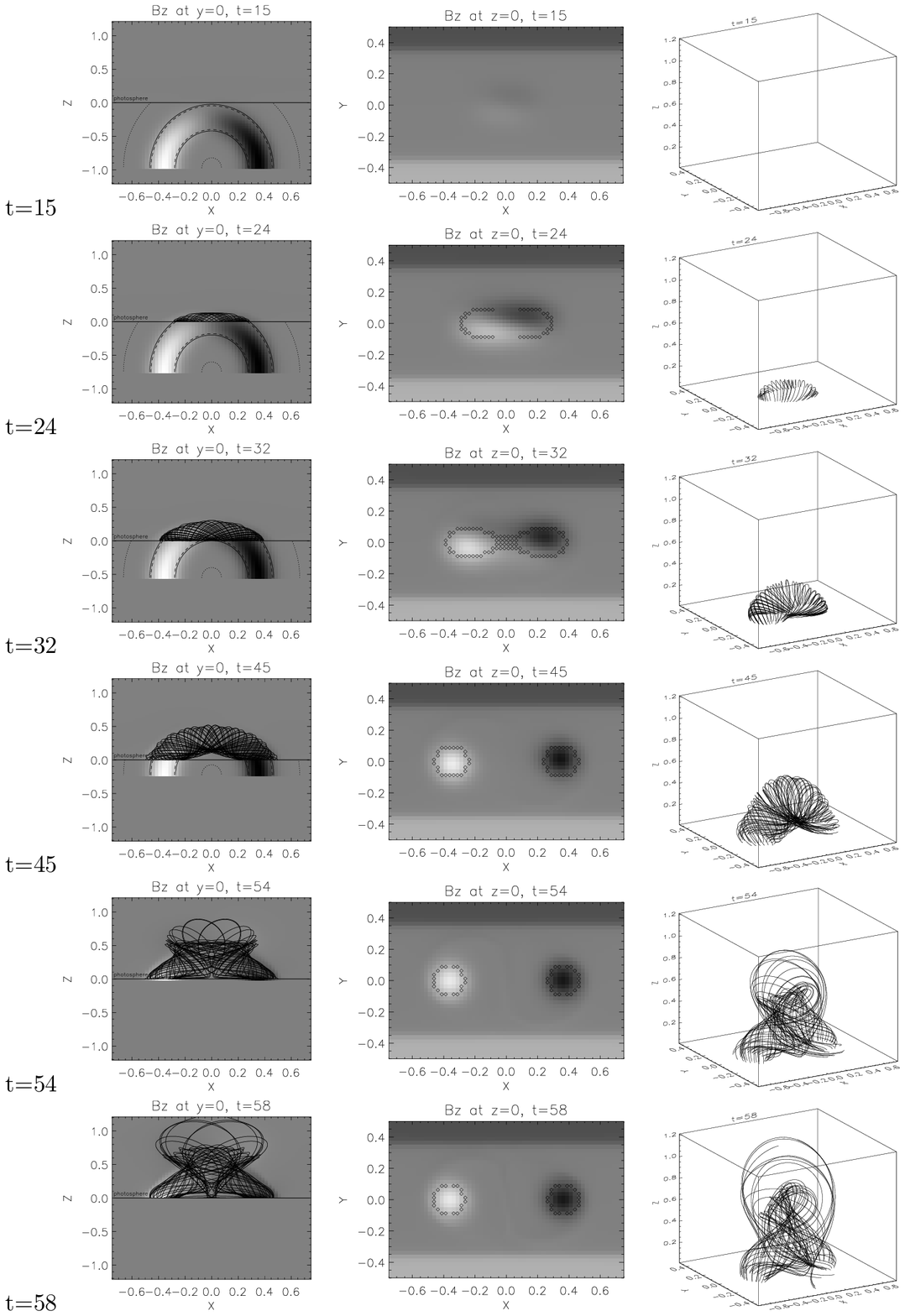} \\
 \end{center} 
  \caption{\small{The characteristic times for the simulation of Fan \& Gibson, 2003,
   different rows correspond to different time (see detailed explanation in text). 
   \textit{First column} -- $XZ$ slices, the analytical shape of the rising 
   tube is shown beyond the photosphere, solid-dashed line is $\varpi=1.0$ 
   -- the formal ``edge'' of the torus; dotted line is $\varpi=3.0$. 
   \textit{Second column} -- magnetograms at $z=0$. 
   \textit{Third column} -- side view of the field lines, initiated at 
    $\varpi=1.0$ (their footpoints are shown as diamonds in the second column).}}
  \label{relative1}
 \end{figure}

A visual representation of characteristics times is shown on
Fig.~\ref{relative1}. Different rows correspond to different
times: $t=15$ -- the tube is about to emerge; $t=24$ -- the minor axis
of the torus has emerged; $t=32$ -- the bottom of the torus has
emerged; $t=45$ -- the tube undergoes acceleration; $t=54$ -- the
major axis of the torus has emerged, the torus has stopped emerging,
the tube starts getting a significant writhe; $t=58$ -- the tube
escapes the domain; the simulation is over. Note that the torus starts
to kink at $t\geq 45$ and keeps kinking until it escapes the
computational domain at $t=58$. \\

  \begin{figure}[!hc]
 \begin{center}
    \includegraphics[width=6cm]{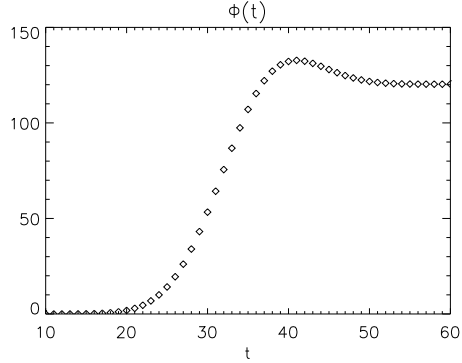} 
 \end{center} 
  \caption{\small{The total amount of unsigned photospheric flux, as defined in Eqn.~(\ref{flux}), of the torus (not counting the arcade), plotted as a function of time. The major axis of the torus emerges at $t=54$. The maximal value of flux is reached earlier than that because of the field winding aroung the torus and thus being not necessarily normal to $\zhat$. After $t=54$ the torus has stopped emerging and thus the magnetic field at the photosphere remains constant. }}
  \label{flux} 
 \end{figure}

\clearpage

\subsection{Computing $H_A$ For Given Volume And The Potential Field.}

We define different domains, $\dmn$, with the same field, by making
a different choices of boundary mask. We were interested in how
different portions of the torus, namely, the ``core'' and the outer
layers behave during the instability. \\ 

\begin{figure}[!hc]
 \begin{center}
    \includegraphics[width=6cm]{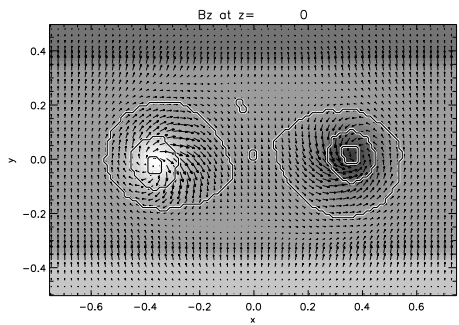} 
 \end{center} 
  \caption{\small{An example of footpoints of domains
$\Theta\left(\varpi_{max}\in\left[0.5, 1.0, 2.0\right],
t=50\right)$. The vertical field, $B_z$, is shown in grayscale and horizontal field is shown with arrows. Three pairs of concentric curves, counting from inside out enclose footpoints of the domains defined by $\varpi_{max}=0.5$, $\varpi_{max}=1.0$ and $\varpi_{max}=2.0$.}}
  \label{footpoints_diff_po} 
 \end{figure}

Our masks are defined to be within the photospheric intersection of the
emerging torus, $\varpi \leq \varpi_{max}$.
By choosing different values of $\varpi_{max}$ we construct
domains, containing different portions of the emerging flux tube 
The footpoints of domains with different
$\varpi_{max}$ are shown on Fig.~\ref{footpoints_diff_po}. The shape
is distorted with respect to the original cross-section of a torus due
to reconnection with the arcade, current sheet formation and due to
near horizontality of some field lines. \\

\begin{figure}[!hc]
 \begin{center}
   \includegraphics{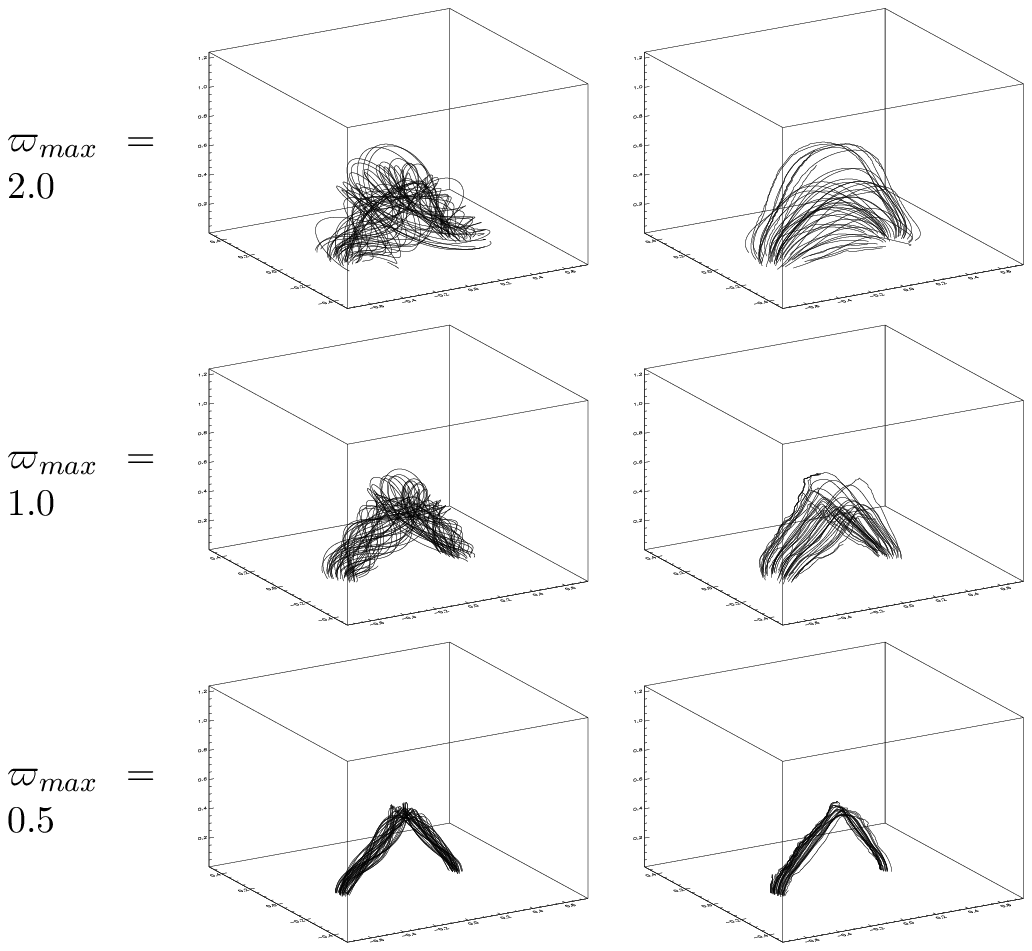} 
 \end{center} 
  \caption{\small{\textit{(left column)} -- field of a torus, 
           confined to domains of different $\varpi_{max}$, 
           with footpoints shown on Fig.~\ref{footpoints_diff_po}.
           \textit{(right column)} -- the potential field, 
           constructed for each such domain. }}
  \label{diff_po_3d}
 \end{figure}

\begin{figure}[!hc]
 \begin{center}
   \includegraphics{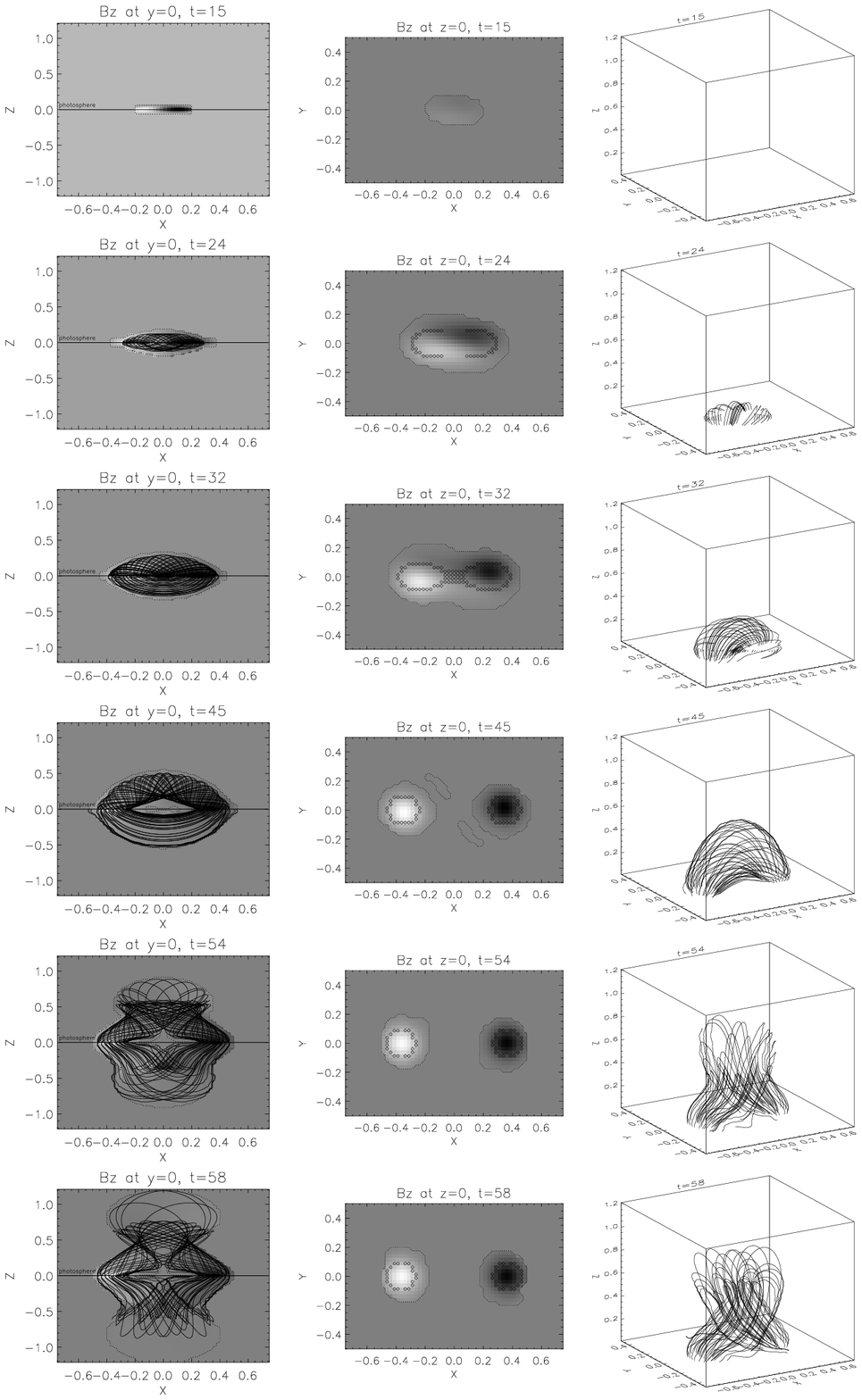} 
 \end{center} 
  \caption{\small{The strip plot of the results of the computation. The original 
   data is shown above and the relaxed potential field $\pvec$ -- 
   below the photosphere. The dotted line indicates slices of the domain $\Theta\left(\varpi_{max}=2.0\right)$. 
   The magnetogram in the second column and the field lines in the third column 
   are those of $\pvec$. All notation is similar to Fig.~\ref{relative1}.}}
  \label{relative2}
 \end{figure}

We found domains for masks with 
$\varpi_{max}\in\left[0.5,1.0,2.0\right]R$ at different
times during the emergence.  
We computed $\Theta\left(\varpi_{max},t\right)$
and then constructed a
potential field confined to it. The results are shown in
Fig.~\ref{footpoints_diff_po}, Fig.~\ref{diff_po_3d} and
Fig.~\ref{relative2}. \\ 
  
\clearpage

\clearpage

For each $t$ and $\varpi_{max}$ we calculated vector potentials of the
actual field,
$\Theta\left(t, \varpi_{max}\right)\bvec\left(\rvec, t\right)$, and
the reference field
$\pvec\left(\rvec, t, \varpi_{max}\right)$. To do this we used a
gauge in which one of the components of the vector potential (in our
case, $A_z$) is identically zero. The other two could be found
with a straight-forward computation: 
\begin{equation}\begin{array}{rcl}
 A_x(x, y, z) & = & \int\limits_{0}^{z}B_y(x, y, z')dz'\\
 A_y(x, y, z) & = & f(x, y)-\int\limits_{0}^{z}B_x(x, y, z')dz'\\
 f(x, y) & = & \int\limits_{0}^{x}B_z(x', y, 0)dx'
\end{array}\end{equation}
In terms of these elements the addirive self helicity
the additive self-helicity: 
\begin{equation}
	H_{A}\left(t, \varpi_{max}\right)=\int\limits_{\Theta\left(t, \varpi_{max}\right)}
	{ \left(\Theta\bvec-\pvec\right)\cdot\left( A+A_P \right) dV }
\end{equation}
is computed.

\subsection{Measuring Twist in Thin Flux Tube Approximation}

To make contact with previous work we compare the additive self
helicity to the twist helicity in our flux bundeles.
It can be shown analytically that in the limit of a vanishingly thin
flux tube these quantities are identical.  Here we must compute twist
helicity for flux bundles of non-vanishing width.  We do this in terms
of a geometrical twist related to twist helicity.

One cannot really speak of twist, or of an axis, in the domains defined
above.  First, the thickness and the curvature radius of the flux
bundles are comparable to their lengths.  Secondly, the 
magnetic field and the twist vary rapidly over the cross section of
the bundle.\\

The domains constructed from the smaller masks, $\varpi_{max}=0.5$
and $\varpi_{max}=1.0$, may, however, be suitable for approximation
as thin tubes.  Even in these cases the
approximation may suffer near the top part at later times: at
$t=50$ the radius of curvature becomes comparable to the width, and later,
during kinking the radius of the tube becomes comparable to the 
length (see Fig.~\ref{footpoints_diff_po} and
Fig.~\ref{relative1}). 

We define an axis for the flux bundle by first tracing many field
lines within it. Then we divide each field line into $N$ equal segments
($N$ is the same for all lines) of length $L_i/N$, where $L_i$ is the
length of the $i^{\rm th}$ line. If the bundle were an ideal cylinder, 
the midpoints of the $n^{\rm th}$ segment from every line 
would lie on a single plane; provided the bundle is thin the
these midpoints will lie close to a plane. We define the $n^{\rm th}$
point on an axis by the centroid of these approximately co-planar points. 
The set of $N$ centroids forms the axis of our tube. \\ 

We then define the tangent vector $\lhat_i$
along this axis, and a plane normal to this vector and thus 
normal to the flux tube (at least in the thin flux tube approximation).
If the tube has some
twist in it, then the point where one field line intersects the plane
will spin about the axis as the plane moves along the tube. Such
spinning must be defined relative to a reference vectore on the plane
which ``does not spin''.  The net angle by whcih the intersection
point spins, relative to the non-spinning vecotr, is the total twist
angle of the tube.  In a thin tube all field lines will spin by the
small angle; in our general case we compute an everage angle.\\

\begin{figure}[!hc]
 \begin{center}
   \includegraphics[angle=0, width=5.0cm]{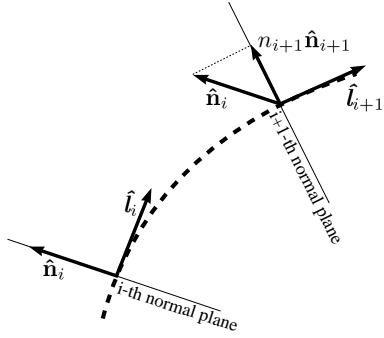}
 \end{center} 
  \caption{\small{An illustration of parallel transport of a coordinate system. At every next step one unit vector of the previous coordinate system is projected to a new normal plane and normalized; the second vector is created anew as perpendicular to the new unit vector. $\lhat$ is the tangent vector of the axis, $\nhat$ is the unit vector in normal plane, carried with the plane along the axis.}}
  \label{parallel_transport}
 \end{figure}

We produce a non-spinning reference vector using an orthonormal triad,
arbitrarily defined at one end of the tube, and carried along the axis
by \textit{parallel transport}.  For a curve with tangent
unit vector $\lhat$, the parallel transport of a vector $\uvec$ means
$\lhat\cdot\left(\partial\uvec/\partial l\right)=0$.  To impliment
this numerically
an arbitrary unit vector $\uhat_0$ is chosen at one end of the
axis perpendicular to the tangent, $\uhat_0\cdot\lhat_0=0$.  The third
member of the triad is
$\vhat_0=\uhat_0\times\lhat_0$. At the next point, $\uhat_1$ is chosen
by projecting $\uhat_0$ onto a plane normal to $\lhat_1$ and
normalizing it
$$\uhat_1=\frac{\uhat_0-\left(\uhat_0\cdot\lhat_1\right)\lhat_1}
{\left|\uhat_0-\left(\uhat_0\cdot\lhat_1\right)\lhat_1\right|},$$
(see Fig.~\ref{parallel_transport}). Then
$\vhat_1=\uhat_1\times\lhat_1$, and the procedure is repeated for
every segment along the axis. 

 \begin{figure}[!hc]
 \begin{center}
   \includegraphics{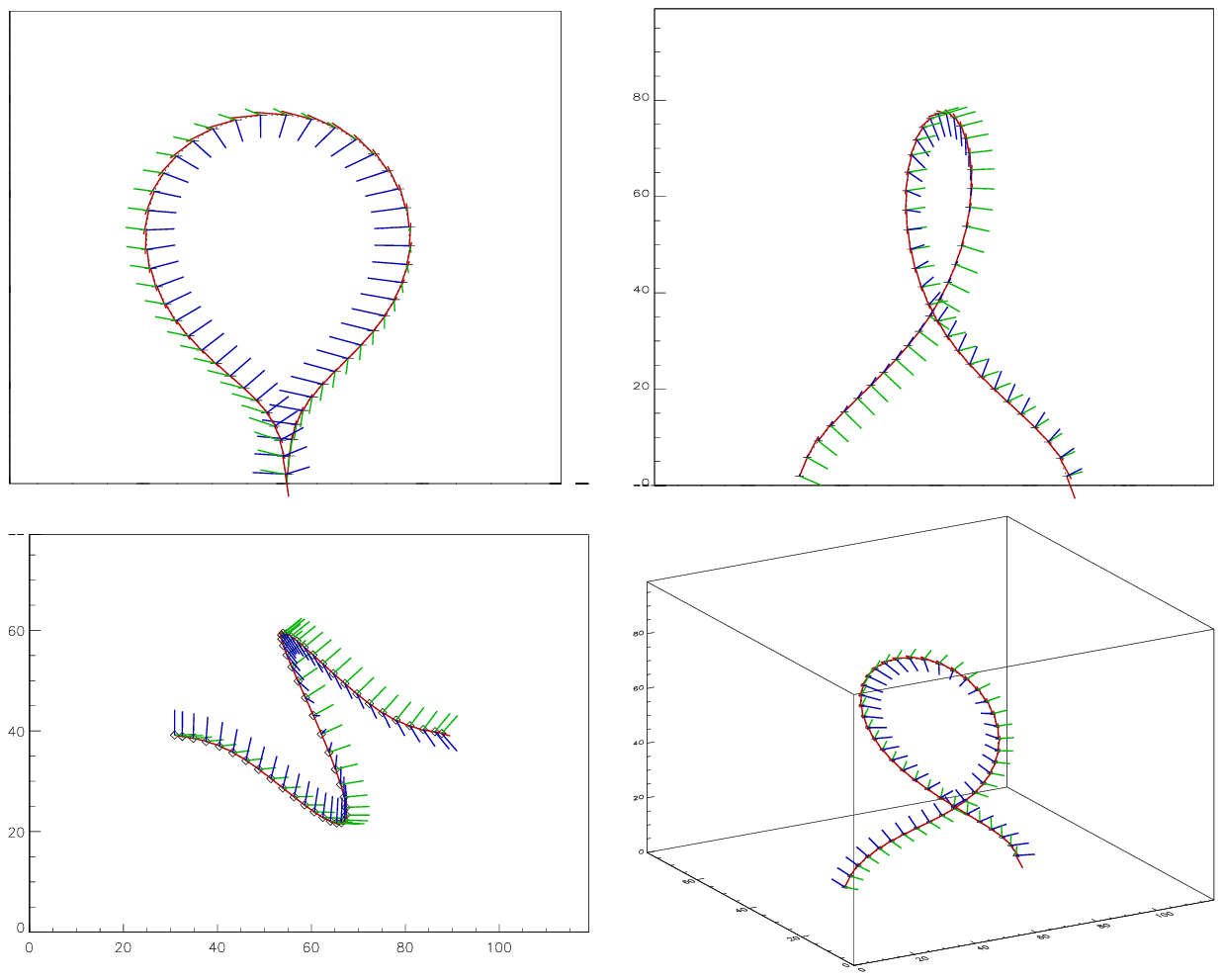}
 \end{center} 
  \caption{\small{An example of the axis, found for $\varpi_{max}=0.5$, $t=58$, and the corresponding coordinate system, carried along by parallel transport. $\lhat$, $\uhat$ and $\vhat$ are drawn in red, green and blue colors respectively.}}
  \label{parallel_transport_1}
 \end{figure}

 \begin{figure}[!hc]
 \begin{center}
   \includegraphics{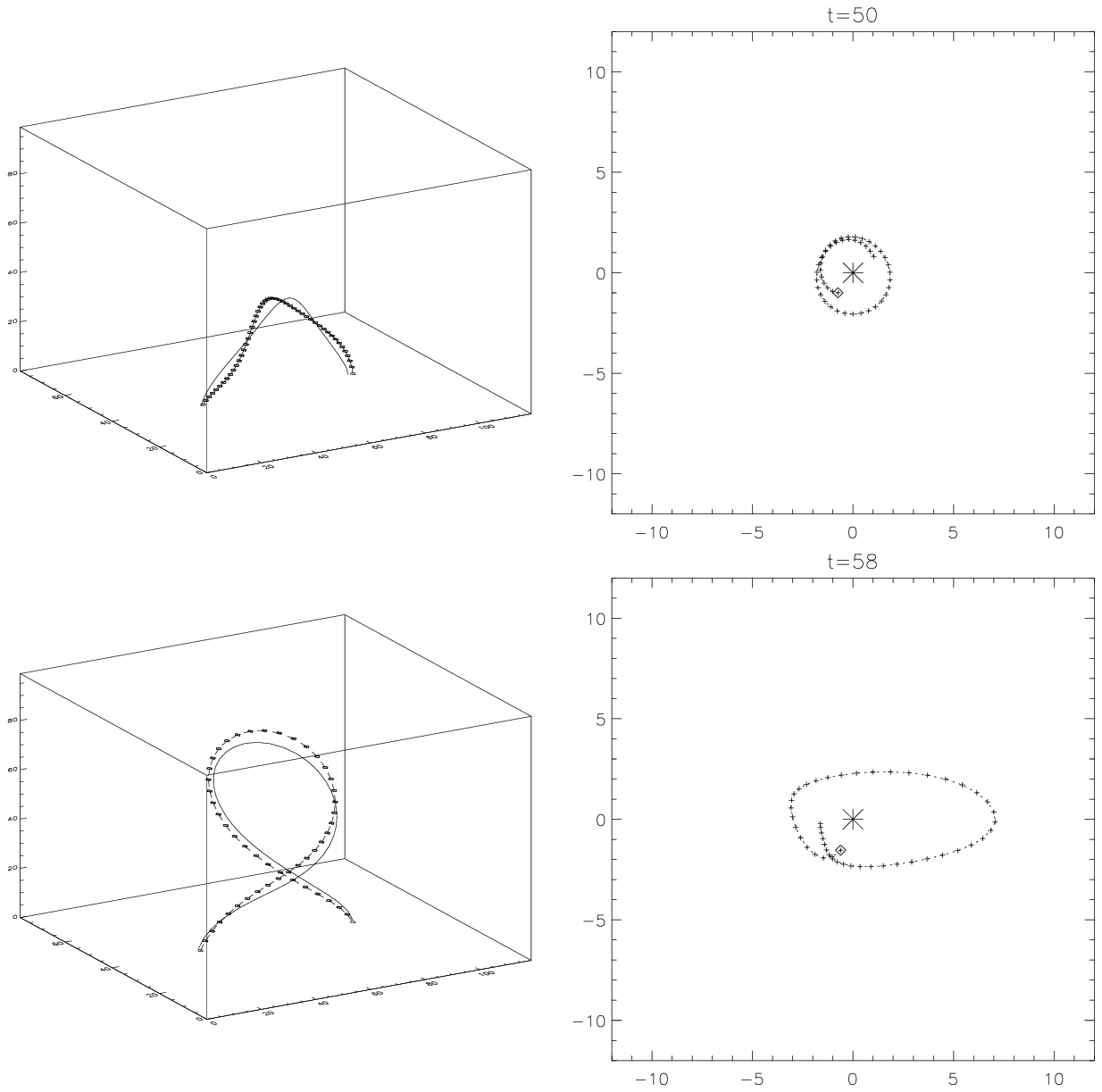}
 \end{center} 
  \caption{\small{An illustration of how kinking decreases twist. 
   An axis (solid) of a ``thin'', $\varpi_{max}=0.5$, tube and a single field line 
   (dotted with diamonds) at a different times: top row is $t=50$, the field line has $\Delta\theta\approx-3.1\pi$ and bottom row is $t=58$ and the field line has $\Delta\theta\approx-2.4\pi$ (and $Tw=\Delta\theta/2\pi$). Left column is sideview and right column is \textit{the trajectory of the line} in the tangent plane with coordinate system decribed above.}}
  \label{honest_tw_singleline}
 \end{figure}


\section{Results}

Based on the analogy between $Tw$ and $H_A/\Phi^2$, it would
be natural to introduce quantity analogous to $L$ and $Wr$ in a
similar way. We propose that $L$ in the general (non-``thin'') case
might be analogous to the \textit{unconfined self-helicity},
introduced in Longcope \& Malanushenko, 2007\nocite{Longcope2008e},
and $Wr$ is similar to the helicity of the confined potential field
relative to the unconfined potential field.
From equation (3) of Longcope \& Malanushenko, 2007\nocite{Longcope2008e} 


\begin{equation}
H(\bvec, \pvec_\vol, \vol)\equiv\int\limits_{\vol}{d^3x\bvec\cdot\avec}-\int\limits_{\vol}{d^3x\pvec_\vol\cdot\avec_P}+\int\limits_{z=0}{dxdyB_z(x, y, 0)\int\limits_{\xvec_0}^{\xvec}{d\xvec'\left[\avec(\xvec')-\avec_P(\xvec')\right]} }
\end{equation} 
(where $\xvec=\rvec(x, y, 0)$ and $\pvec_\vol$ is a potential field confined to $\vol$ that matches 
boundary conditions $\bvec\cdot\nhat|_{\partial\vol}=\pvec_\vol\cdot\nhat|_{\partial\vol}$) 
by plugging it into $H(\Theta_\dmn\bvec, \pvec_\dmn, \vol)$ and $H(\bvec, \pvec_{\vol, \Theta}, \vol)$ 
and adding them together it immediately follows, that 
\begin{equation}H\left(\Theta_\dmn\bvec, \pvec_\dmn, \vol \right)+
H\left(\pvec_\dmn, \pvec_{\vol, \Theta}, \vol \right)=
H\left(\Theta_\dmn\bvec, \pvec_{\vol, \Theta}, \vol \right), 
\end{equation}
where $\dmn\subset\vol$ and $\Theta_\dmn$ is a support function of $\dmn$.  
By $\pvec_\dmn$ we mean the potential field confined to $\dmn$ (and 
identically zero outside of $\dmn$) that matches boundary conditions 
$\bvec\cdot\nhat|_{\partial\dmn}=\pvec_\dmn\cdot\nhat|_{\partial\dmn}$, 
and by $\pvec_{\vol, \Theta}$ we mean the potential field, confined to $\vol$, that 
matches boundary conditions 
$\Theta_\dmn\bvec\cdot\nhat|_{\partial\vol}=\pvec_{\vol, \Theta}\cdot\nhat|_{\partial\vol}$.
As long as $\dmn$ is fully contained in $\vol$, which is constant in
time, the quantity $H_{\rm unc, \vol}/\Phi^2\equiv H\left(\Theta_\dmn\bvec,
\pvec_{\vol, \Theta}, \vol \right)/\Phi^2$ will \textit{behave
like} $L$ and $H\left(\pvec_\dmn, \pvec_{\vol, \Theta}, \vol \right)/\Phi^2$ 
would then \textit{behave like} $Wr$. \\

\begin{figure}[!hc]
 \begin{center}
    \includegraphics[width=10cm]{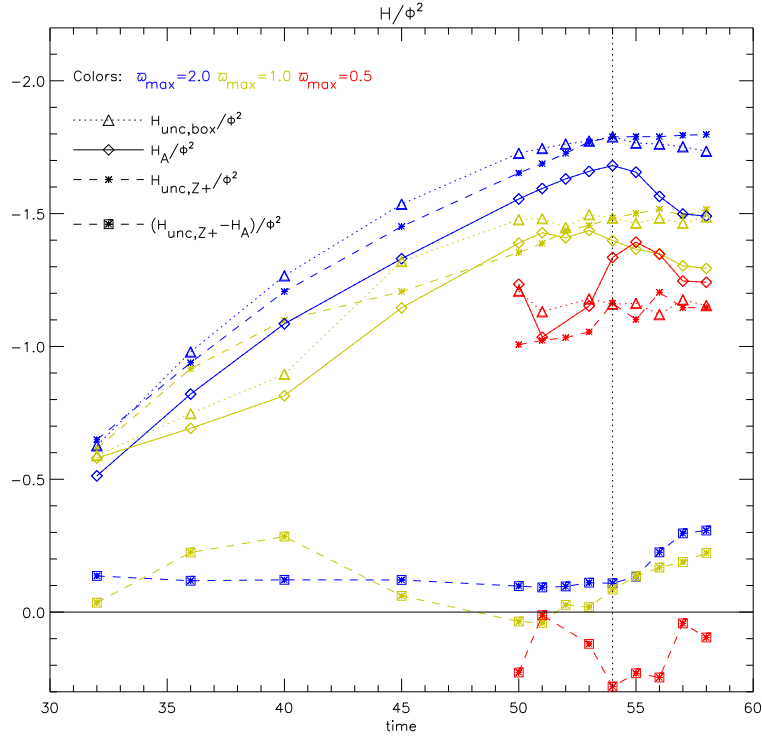} 
 \end{center} 
  \caption{\small{The comparison between $H_A$ (i.e., confined to the
volume of the flux tube), $H_{\rm unc, box}$ (confined to the box in which the
original simulation was performed) and $H_{\rm unc, Z_+}$ (confined to
half-space), normalized by $\Phi^2$. Vertical dashed line at $t=54$
indicates the time when the emergence has stopped and \textit{all}
further changes in $Tw$ would be due to kinking and numerical
diffusion, and all earlier changes are altered by the emergence of the
tube and thus non-zero helicity flux over the surface. For $\varpi_{max}$ of
$2.0$ and $1.0$ it's clearly visible, that: a) after $t=54$ the
unconfined helicities remain nearly constant, while the confined to
flux bundle's $\dmn$, that is, additive self helicity, decreases due
to kinking; b) before $t=54$ the difference between $H_{\rm unc, Z_+}$, that
is, the integrated helicity flux, and $H_{\rm unc, box}$ is non-zero. 
The threshold for $H_A/\Phi^2$ seems to be $-1.7$ for
$\varpi_{max}=2.0$ and $-1.4$ for $\varpi_{max}=1.0$. $\varpi_{max}=0.5$ seems to be too
noisy to draw a reliable conclusions; possible reasons for that are
discussed in the text.}}
  \label{unc_vs_hel} 
 \end{figure}

Fig.~\ref{unc_vs_hel} compares 
the generalized twist number, $H_A/\Phi^2$, with helicity, unconfined to the 
flux bundle's volume, but confined to the computational domain of the simulation: 
$H_{\rm unc, box}$. In this case $\vol$ is the
computational domain, a rectangular box.   The behaviour of all quantities matches expecation: 
$H_{\rm unc, box}/\Phi^2$ 
increases as the torus emerges, and stays nearly constant after the
emergence is complete (the slight decrease is due to the reconnection
with the arcade field). The generalized twist number, 
$H_{A}/\Phi^2$ also increases with the
emergence, but decreases between $t=50$ and $t=58$ -- the time when
the torus kinks (see Fig.~\ref{relative1}).   For different
$\varpi_{max}$ the decrease seems to start at a slightly different
time.  
\\

Fig.~\ref{unc_vs_hel} demonstrates as well, that the general behaviour
of $H_{\rm unc, \vol}/\Phi^2$ is qualitatively similar whether the
volume $\vol$ over which unconfined helicity is computed is the
computational domain or the half space. To compute the unconfined
helicity in the half space, $H_{\rm unc, Z_+}$, we integrate the helicity flux in the way
described in \citep{devore2000} and used in \citep{fan2004}. The
helicity flux is computed relative to the potential field in half
space, and thus, the helicity flux, obtained in this way, might be
considered a ``confined to a half space''. 

Longcope and Malanushenko (2008) show that $H_{\rm unc, box}=H_{\rm  
unc, Z_+}$ when the volumes, $\vol$ and $Z_+$ and the vertical field, $B_z(z=0)$, all  
share a reflectional symmetry. This situation occurs in the simulation only for $t\geq 54$ when the  
torus is fully emerged and its major axis is at the photosphere.  At these times the vertical  
component of the field is the toroidal component of the torus, which is symmetric about $y=0$.  Due  
to reconnection with the arcade, however, the footpoints of $\dmn$ may not share this symmetry, in which case  
the photospheric field $\Theta_\dmn B_z$ is not precisely symmetric.  If  
the two helicities were ever to coincide, it would be at $t=54$, so  
we choose constant of integration by  setting $H_{\rm unc, box}=H_ 
{\rm unc, Z_+}$ at that time.  The time histories of both unconfined  
helicities are plotted in Fig.~\ref{unc_vs_hel}. The discrepancy  
between the two before $t=54$ arises from the non-vanishing  helicity  
of $\pvec_{\vol,\Theta}$ relative $\pvec_{Z_+,\Theta}$ owing to a photospheric  
field, $B_z$, lacking reflectional symmetry. In spite of the  
discrepancy, we draw from each curve the same basic conclusion, that  
the kink deformation of $\dmn$ does not change $H_{unc, \vol}$.

\begin{figure}[!hc]
 \begin{center}
    \includegraphics[width=10cm]{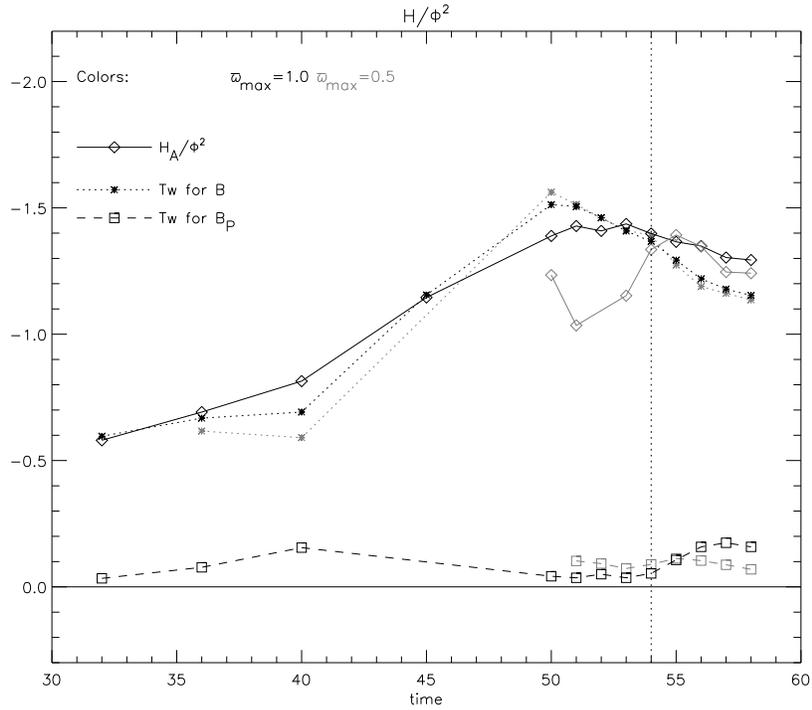} 
 \end{center} 
  \caption{\small{A comparison between generalized twist number (solid
line with diamonds) and the ``thin tube'' classical twist number
(dotted line with asterisks) for two subvolumes of a different
size. Also, the ``classical'' twist number for a potential field
(dashed line with squares).}} 
  \label{tw_vs_hel} 
 \end{figure}

Fig.~\ref{tw_vs_hel} compares the generalized twist number to the
traditional twist number described above. The twist number was 
computed only for the thinner subvolumes of the torus,
$\varpi_{max}=0.5R$ nd $\varpi_{max}=R$.
Fig.~\ref{tw_vs_hel} shows agreement quite well for
$\varpi_{max}=R$ and less well for $\varpi_{max}=0.5R$. 
The reason might be the following: the smaller the subvolume, the 
the fewer points does it have, so that, first, there are fewer field lines 
to be traced to measure twist, and second, the potential field, obtained by 
relaxation is numerically less precise. Nevertheless, the magnitudes 
and the general behaviors do agree.

Fig.~\ref{tw_vs_hel} also shows the twist number measured for
the potential field in a subvolume $\pvec$, is zero to measurement
error. Note, that a significant portion of the torus is emerged, its
length is not large enough (relative to the thickness) for the
thin tube approximation to be valid. As the twist of the potential
field should theoretically be zero (as well as generalized twist),
this plot also gives an idea of the magnitude of the error of
twist measurements; at most times the error is less than 15\% of the value. \\ 

\clearpage
\section{Discussion}

We have demonstrated that, at least in one MHD simulation, the quantity, 
$Tw_{(gen)}$, defined in terms of the additive self helicity shows a threshold 
beyond which the system became dynamically  
unstable.  The simulation we considered, originally studied by 
\citet{Fan2003}, is a three-dimensional, numerical solution of the time- 
dependent, non-linear evolution of an emerging flux system.  The  
original study established that the system became unstable to a  
current-driven (kink) mode at some point during its evolution.  In  
this work we have shown that the quantity $Tw_{(gen)}$ increases  
until the instability ($Tw_{(gen)}\simeq 1.5$) at which time it  
drops.  This drop occurs as a natural consequence of the instability  
itself.

The quantity we propose as having a threshold,  $Tw_{(gen)}$, is  
computed using a version of the self helicity previous defined by
\citet{Longcope2008e}.  The present work has provided a detailed  
method for computing this quantity for any complex bundle of field  
lines within a magnetic field known on a computational grid.  We also  
demonstrate that for the very special cases when that bundle can be  
approximated as a thin flux tube, $Tw_{(gen)}$ is approximately equal to the  
traditional twist number, $Tw$.  In the case of thin flux tubes which  
are also dynamically isolated, free magnetic energy is  
proportional to $(Tw)^2$. Their free energy may be spontaneously  
reduced if and when it becomes possible to reduce the magnitude of 
$Tw$ at the expense of the writhe number, $Wr$, of the tube's axis.

All this supports the hypothesis that $Tw_{(gen)}$ could be treated  
as a generalization of $Tw$.  Such a generalization might be extremely  
useful
in predicting the stability of magnetic equilibria sufficiently  
complex that they cannot be approximated as thin flux tubes. The case  
we studied, of a thick, twisted torus of field lines \citep{Fan2003},  
appears to become unstable when $Tw_{(gen)}$ exceeds a threshold  
value between $1.4$ and $1.7$. This value happens to be similar to the threshold 
on $Tw$ for uniformly twisted, force-free flux  
tubes, $Tw\approx1.6$, as $\Delta\theta\approx3.3\pi$ \citep{Hood1979}.

Previous investigations have shown that the threshold on $Tw$ depends  
on details of the equilibrium such as internal current distribution  
\citep{Hood1979}.   It is reasonable to expect the same kind of  
dependance for any threshold on $Tw_{(gen)}$, so we cannot claim that  
$Tw_{(gen)}<1.7$ for all stable magnetic field configurations.  
To investigate such a claim is probably intractable, but useful insights 
may be obtained by applying the above analysis to magnetic equilibria 
whose stability to the current-driven instability is already known.
The paucity of closed-form, three-dimensional equilibria in the literature, 
and far fewer stability analyses of them, suggests this may be 
a substantial undertaking.

\bibliography{c:/localtexmf/bib/short_abbrevs,c:/localtexmf/bib/full_lib,c:/localtexmf/bib/my_bib}
\end{document}